\documentstyle[psfig]{lamuphys}
\makeatletter
\let\chapter\hid@chapter
\makeatother

\begin{document}
\pagenumbering{arabic}

\title{Searches for High Redshift Clusters}

\author{Mark Dickinson}

\institute{STScI, 3700 San Martin Dr., Baltimore MD 21218 USA}

\maketitle

\begin{abstract}
High redshift galaxy clusters have traditionally been a fruitful place
to study galaxy evolution.  I review various search strategies for finding
clusters at $z > 1$.  Most efforts to date have concentrated on the 
environments of distant AGN.  I illustrate these with data on the cluster
around 3C~324 ($z=1.2$) and other, more distant systems, and discuss
possibilities for future surveys with large telescopes.
\end{abstract}

\section{Finding distant clusters}
At one time, galaxy clusters served as the most observationally straightforward
means of studying galaxy evolution at high redshift.  The reason was primarily
one of {\it contrast}:  even without spectroscopy, very rich clusters are 
recognizable as enhancements in the galaxy surface density out to $z \sim 1$,
and the properties of the cluster galaxy population can be studied
statistically with imaging data alone if a proper ``control sample'' of 
field galaxies can be observed in the same manner.  In this way, Butcher
and Oemler (1978, 1984) provided the first convincing evidence for galaxy 
evolution, identifying an apparently systematic bluing trend for galaxies 
in the cores of rich clusters beyond $z \approx 0.3$.  

At very large redshifts ($z > 1$), however, even quite rich clusters are 
no longer so clearly visible against the tremendously numerous population of 
faint field galaxies, and thus become correspondingly more difficult to
discover and study.  This is illustrated by figure 1, a cartoon representing
the peak surface density contrast in the $I$--band of identical rich clusters
observed at $z = 0.3$ and $z = 1.2$.  Because the angular scale changes only
slowly with redshift beyond $z = 0.3$, there is very little boost in the
galaxy surface density as one moves a cluster further away, while the 
combination of distance modulus and k--correction dims the cluster galaxies
considerably.  The result is that a cluster with a contrast $10\times$
over the field galaxy population at $z = 0.3$ barely peaks up over the
field at all at $z = 1.2$.  The effects of the k--correction are much more
severe for early--type galaxies, rendering a rich, elliptical dominated
cluster nearly invisible at optical wavelengths for $z > 1$, even assuming
reasonable amounts of passive evolution.

\begin{figure}
\centerline{\psfig{figure=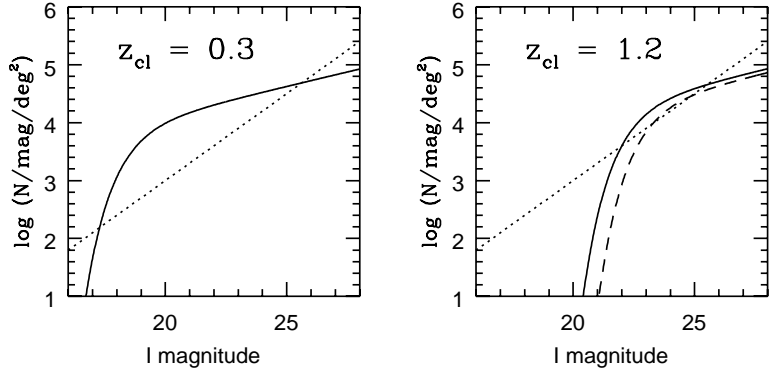,width=5in}}
\caption{Schematic plot of surface density contrast for a rich cluster
observed at $z=0.3$ and 1.2.  The dotted line sketches $I$--band field galaxy 
counts, while the solid curve represents the central galaxy surface density 
of a rich cluster with a Schecter luminosity function.  At $z=0.3$, the 
cluster core has a peak contrast of $\sim 10\times$ the background, while at 
$z=1.2$ the contrast is $< 2\times$.  A k--correction for Sb--type galaxies 
with mild luminosity evolution has been assumed -- the contrast is reduced to
virtually nothing if the cluster galaxies are all ellipticals (dashed line) 
due to the stronger k--correction.}
\end{figure}

On the one hand, thanks to highly efficient multiplexing spectrographs,
the tremendous progress which has occurred in the study of high redshift field 
galaxies has meant that clusters are no longer ``needed'' to provide large 
samples of distant galaxies.  But at the same time, the higher the 
redshift, the more interesting a rich cluster becomes
from a cosmological viewpoint.  Firstly, as the most massive collapsed 
structures in the universe, their properties and evolution are highly 
sensitive to the fundamental cosmological parameters, as well as to the
power spectrum of mass fluctuations which give rise to large scale structure
in the galaxy distribution.  The abundance of massive clusters, and their 
detailed properties, become increasingly important constraints on 
cosmological models at larger redshift.  Secondly, today's rich clusters
are dominated by elliptical galaxies, whose evolutionary history is of
particular interest since they may represent the oldest galaxy--sized
stellar systems.  Tracing their spectrophotometric properties may point
back to the earliest epochs of star formation.

The observational challenge in finding and studying distant clusters is
one of enhancing their {\it contrast} against the field.  Simply imaging
in a single, broad bandpass is not sufficient.  In the absence of extensive
spectroscopy, multi--color or narrow band techniques must be used to
screen out the multitude of mostly foreground galaxies and to isolate candidate
cluster members.  Here, I consider a variety of means by which this 
may be achieved.  Many of these are costly in terms of observing 
time:   regardless of the method, one must work to very faint 
apparent magnitudes in order to see distant clusters, requiring large
telescope apertures and long exposure times.  Thus far, virtually all
attempts to find extremely high--$z$ clusters have been ``targeted'' surveys.
Rather than blindly search the sky for contrast enhancements in the galaxy
density, targeted searches select likely sites of distant clusters and
concentrate their efforts there.  Because galaxies cluster, if one already 
knows where one high--$z$ object is, then that is a good place to start looking
for more.  High redshift AGN are thus a natural place to begin searching
for fainter, more normal companions.  In particular, various previous
surveys (e.g. Yee \& Green 1984, Yates {\it et al.} 1989, Hill \& Lilly 1991)
have found that radio--loud AGN at 
$z \approx 0.5$ are frequently (but not always) situated in rich galaxy 
clusters.  For this reason, my collaborators and I, as well as others,
are studying the environments of distant radio galaxies and quasars.
Our initial results have been encouraging, and suggest that this may
be a fruitful pursuit for future surveys with large telescopes such as 
the VLT.

\section{A case study:  3C 324}

One means of enhancing the contrast of distant clusters is to search in
the near--infrared, where the field galaxy number counts exhibit a shallower
slope, and where the k--correction which dims the light of distant early--type
galaxies is significantly reduced.  Peter Eisenhardt and I have been carrying
out a systematic survey of the environments of radio galaxies at 
$0.8 < z < 1.4$, using deep infrared and optical imaging to search for
enhancements in the galaxy surface density around the AGN.  Several
promising cluster candidates have been found in this manner.  
In order to confirm these and to study their properties in more detail, 
Hy Spinrad, Arjun Dey and I have been following these up with Keck multislit
spectroscopy.  We have also obtained deep {\it HST} images of two of these
clusters so far.  Here I will use the environment of 3C~324, a 
powerful radio galaxy at $z=1.206$, to illustrate the challenges and
promise of these methods.

\begin{figure}
\centerline{\psfig{figure=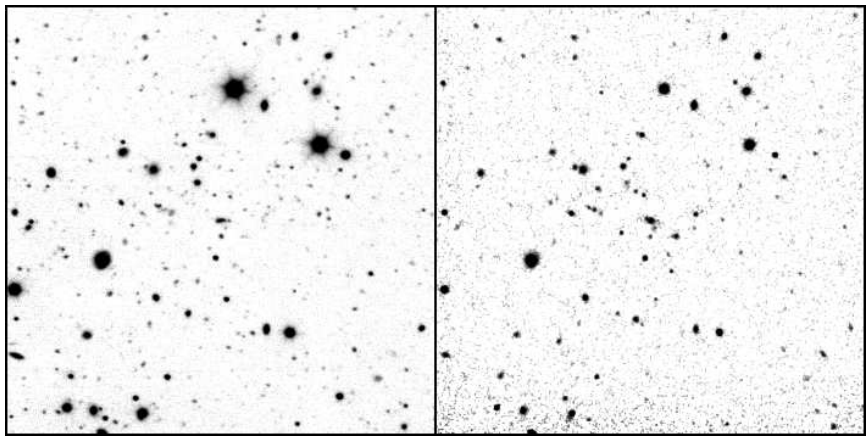,width=4.5in}}
\caption{Deep $R$ (left) and $K$ (right) images of a $2^{\prime}.5$ field 
around the $z=1.206$ radio galaxy 3C~324, which sits at field center.  
The cluster is evident in the infrared image, but is nearly invisible in
the optical data.}
\end{figure}

Figure~2 shows deep $R$ and $K$--band images of the field around 3C~324.
To the sharp eye, the cluster appears as a density enhancement in the 
infrared data, but is nearly invisible in the optical image.  Figure~3 
demonstrates this by showing the radial surface density profile
of the cluster in the two images.  Optically, the cluster contrast peaks 
at a factor of only $\sim 2\times$ the background density, whereas in the
infrared the contrast reaches a factor of $\sim 6\times$.  This is because
there are many very red galaxies in this field, which mostly have the expected 
elliptical morphologies in our deep {\it HST} images 
(cf. Dickinson 1995$a,b$).  The k--correction for
these ellipticals is killing the cluster contrast optically, but leaving 
it more prominently visible in the near infrared.   Figure~4 demonstrates
another means by which distant clusters may be recognized:  {\it color
contrast,} wherein a judiciously selected set of broad band filters allows
one to distinguish high--$z$ cluster galaxies from the lower redshift field 
objects by their unusual colors.   

\begin{figure}
\centerline{\psfig{figure=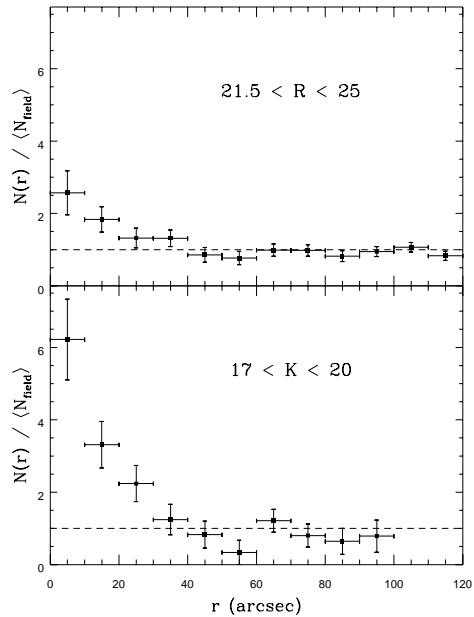,height=3.0in}}
\caption{Radial cluster contrast profile for 3C~324 in the optical and 
infrared, normalized to the ``field'' density of foreground/background
galaxies.  As is evident in figure~2, the cluster contrast is much stronger 
in the $K$--band.}
\end{figure}

\begin{figure}
\centerline{\psfig{figure=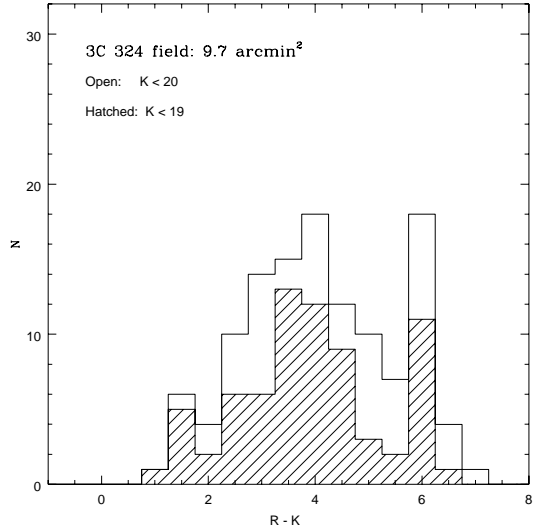,height=3.0in}}
\caption{Histogram of $R-K$ colors for galaxies around 3C~324.  The prominent
spike at $R-K = 5.9$ largely consists of faint galaxies with elliptical
morphologies in the {\it HST} images.}
\end{figure}

Figure~5 shows the results of our spectroscopy around 3C~324, primarily carried 
out using the Keck LRIS, but with additional redshifts provided by Olivier
LeF\`evre from the NTT.  The vast majority of the galaxies we have observed,
particularly at larger separations from the radio galaxy, lie in the foreground 
or background.  However, two sharp spikes in the redshift distribution
appear at $z \approx 1.15$ and $z \approx 1.21$.  These galaxies are strongly
clustered around the radio galaxy:  considering only galaxies at radii
$< 60^{\prime\prime}$ from the radio galaxy, these high--$z$ spikes dominate 
the redshift histogram.  Evidently, the ``cluster'' which appears in the 
$K$--band images is comprised of two distinct clumps or sheets of galaxies 
separated by $\sim 7500$~km~s$^{-1}$ in their rest frame.  Whether this is 
merely a chance projection, or an indication of supercluster--scale structure
at $z \sim 1.2$, is not yet clear.  

\begin{figure}
\centerline{\psfig{figure=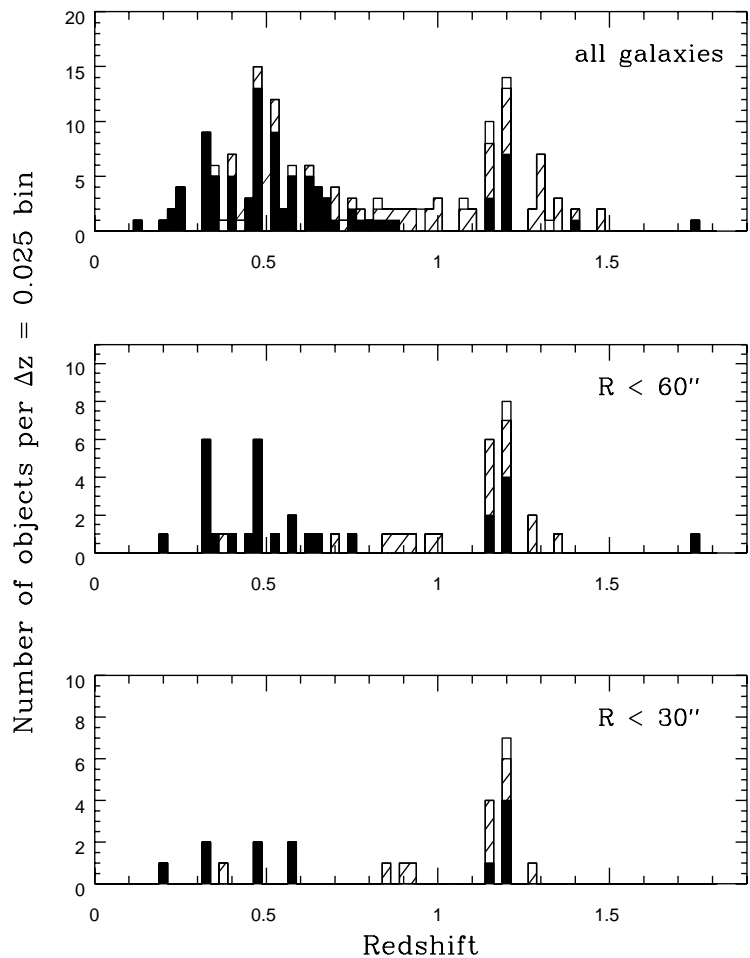,height=5.5in}}
\caption{Redshift histograms of galaxies in the 3C~324 field for samples
restricted at several radii from the radio galaxy.  The various shadings 
indicate different redshift ``quality classes,'' with the filled area 
indicating fully secure redshifts (multiple spectral features) while the 
hatched area largely represents redshifts based on single emission lines,
generally assumed to be [OII].  The ``cluster'' separates into two structures 
at $z=1.15$ and $z=1.21$.  Within $30^{\prime\prime}$, where extended x--ray 
emission is found in our Rosat data, these high--$z$ spikes dominate the 
redshift distribution, suggesting strongly that the x--rays originate at 
$z=1.2$.}
\end{figure}

We believe that one of these two redshift ``spikes,'' probably the one 
associated with the radio galaxy itself, is genuinely a rich, bound cluster.  
As part of an x--ray survey of distant radio galaxies, we obtained a Rosat 
PSPC observation of 3C~324 which detected a faint ($6\sigma$) source within 
$10^{\prime\prime}$ of the radio galaxy position.  At low 
signal--to--noise, the PSPC resolution was insufficient to show whether 
the source is resolved, so the question remained whether or not the
x--rays arose in the radio galaxy AGN or from the cluster environment.
A subsequent 72.1 ksec Rosat HRI exposure clearly shows that the x--ray
emission is resolved over a detectable diameter of $\sim 60^{\prime\prime}$.
The redshift distribution shown in figure~4 exhibits spikes at $z < 1$ 
such as are seen in all faint field galaxy surveys 
(cf. Cohen {\it et al.} 1996).  However, none of the foreground spikes consists
of galaxies particularly concentrated toward the radio galaxy position.
Given the close positional agreement between the x--ray source and 3C~324, 
we regard it as unlikely that the x--rays arise from a foreground group 
or cluster.  If the x--rays indeed originate at $z=1.2$, the 3C~324 cluster has
a bolometric x--ray luminosity $L_X = (8.0 \pm 1.6) \times 10^{44}$~erg~s$^{-1}$
(for $H_0 = 50$, $q_0 = 0.5$), comparable to that of the Coma cluster.
The presence of a large mass is also supported by the detection of weak shear
gravitational lensing centered on the radio galaxy in our WFPC2 images
(Smail \& Dickinson 1995).  This demonstrates the promise of cluster--hunting 
at high redshift:  the properties of such massive, collapsed structures
at $z > 1$ may provide useful constraints on theories of structure
formation and evolution.

\section{Search techniques at higher redshift}

The preceding section demonstrates two methods of contrast enhancement:  
searches in the infrared, and searches by color contrast.  At higher redshifts,
it is unlikely that contrast in a single broad bandpass, even in the 
infrared, will be sufficient to allow recognition of even rich clusters
except perhaps in a statistical fashion.  More carefully ``tuned'' methods
must be adopted.

Just as optical--infrared color contrast helped us to discover the 3C~324
cluster, various {\it infrared--infrared} color combinations may be effective 
for isolating galaxies at $z > 2$.  In particular, combined $JHK$ imaging may 
be effective for detecting galaxies with evolved (ages $>$ 1~Gyr) stellar 
populations at $z ~ 2.5$, where the $J$ band lies shortward of 4000\AA\ in the 
rest frame and the $H$ and $K$ bands roughly measure the rest--frame $B-R$ 
color.  Modelling suggests that older galaxies at $z \sim 2.5$ should have 
identifiably unique colors in this bandpass combination, separating out from 
the locus of lower redshift objects.   

At $z > 3$, the 912\AA\ Lyman limit passes through the observed $U$--band.
This fact has been exploited to great effect by Steidel and collaborators
(cf. Steidel {\it et al.} 1995 and 1996, and the 
contributions of Giavalisco and Macchetto to this conference), who have 
identified and spectroscopically confirmed large numbers of $3 < z < 3.5$ 
field galaxies by selecting them according to their colors in a specially 
tuned $UGR$ filter system.  This is another means of multicolor selection, 
using a spectral break in the rest--frame
UV rather than the rest--frame optical.  Giavalisco (priv. comm.) has used 
this filter system to identify a density enhancement around the $z=3.6$ radio 
galaxy 1243+036, and Lacy \& Rawlings (1996) have reported similar results 
for 4C~41.17 at $z=3.8$.

The Lyman--break technique has been especially effective for studying field
galaxies because it probes a large redshift interval, and hence a large
volume of space.  For this reason, however, it may actually
be less optimal for cluster surveys, where one would ideally like to 
{\it restrict} as much as possible the redshift range over which one isolates 
candidate cluster galaxies.  Narrow band techniques searching for line
emission may therefore also prove useful and effective, although they are
limited to the detection of star--forming galaxies and AGN.  But this may
not be a strong drawback at high redshift, where many or most galaxies
may be actively undergoing star formation.  Lyman~$\alpha$, while optically
convenient for $z > 2$, is a fragile line to work with and is easily
extinguished by dust.  Nevertheless, it has been successfully used by
Francis {\it et al.} 1996 (and this volume), Pascarelle {\it et al.} 1996, 
and M\o ller \& Warren (1993) to identify galaxies in candidate groups or 
clusters at $2 < z < 3$.  LeF\`evre {\it et al.} (see also contribution
by J.--M. Deltorn at this meeting) have also detected and confirmed two 
Ly$\alpha$ companions to the $z = 3.14$ radio galaxy MGO~0316-257.

In the future, narrow band infrared searches may be particularly effective, 
probing Balmer and forbidden line emission which is less affected by dust 
than is Ly$\alpha$ (cf. contributions by Mannucci to this meeting).  
A particularly ``magic'' redshift is $z \sim 2.3$, where [OII], 
[OIII]/H$\beta$, and H$\alpha$ are shifted into the $J$, $H$ and $K$ windows, 
respectively, and where Ly$\alpha$ is shifted to $\lambda_{\rm obs} > 4000$\AA,
facilitating narrow band searches and spectroscopic confirmation.   
At $z \sim 2.3$, the multicolor IR broad band techniques suggested above 
may also be used to look for older, non--star--forming galaxies.  
Indeed, Paul Francis has taken advantage of all of these methods in 
studying the $z=2.38$ system he described in his contribution to this 
conference.

\section{Future prospects and the VLT}

At present, there is very little known about any of these very distant cluster 
candidates except that a few galaxies are present at similar redshifts.  But 
these are early days yet, and preliminary detections can, with intensive 
follow--up studies, lead to more far--reaching results.  For 3C~324, the 
detection of x--ray emission and gravitational lensing gives a first hint at 
cluster {\it masses} beyond $z = 1$.  Our redshift survey of the 3C~324 field,
particularly the unexpected discovery that the ``cluster'' divides into two 
distinct redshift--space structures seen in projection, serves as a reminder
that extensive spectroscopy is needed to confirm and interpret 
any individual cluster candidate.   This is where telescopes like the 
VLT will excell, providing the firepower needed to do this efficiently.

For future cluster surveys, the techniques described above will require
{\it wide field} imaging to very faint flux levels in bandpasses ranging
from $U$ through $K$.  The wide field imaging aspect must be stressed, 
especially in this era of increased angular resolution and adaptive optics, 
which often drives instrument design toward smaller pixels and smaller fields 
of view.   Systems capable of multiplexed imaging (simultaneously observing 
through several bandpasses, split by dichroics) would be highly desirable.
Narrow band capabilities, ideally tunable to any wavelength, are also
likely to be particularly useful for cluster work where restricted redshift
coverage is desirable, much more so than for field galaxy surveys.
Narrow band work is often difficult on very large telescopes because of
the large sizes of the optical beams, but should be considered carefully
in future instrument designs for the VLT.

For detailed follow--up studies of high--$z$ clusters, spectroscopy is the
key.  Proposed future VLT instruments such as VIRMOS will be ideal, 
permitting simultaneous spectroscopy of hundreds of faint galaxies, and
extending into the near--IR where [OII], [OIII] and Balmer emission will be 
redshifted.  And while emission lines are useful and important, the 
success of Steidel {\it et al.} in measuring redshifts from {\it UV absorption 
line} in young, star forming galaxies at $z > 3$ should be kept in mind.  
In the future, infrared continuum and absorption line spectroscopy may 
be essential for studying the properties of {\it older} stellar populations 
of $z > 1$ galaxies in clusters and in the field.

\section{Acknowledgements}

I would like to thank my collaborators, particularly Hy Spinrad, Arjun Dey,
Peter Eisenhardt and Richard Mushotzky, for permitting me to show data in 
advance of publication.  I also thank the conference organizers for an
invigorating meeting in (admittedly frigid) Garching, and for their 
financial support.

\end{document}